\documentclass[letter]{IEEEtran}
\usepackage{cite}
\usepackage{amssymb,amsfonts}
\usepackage[tbtags]{amsmath}
\usepackage{graphicx}
\usepackage{subcaption}
\usepackage{textcomp}
\usepackage{xcolor}
\usepackage{soul}
\usepackage{url}
\usepackage{stfloats}
\usepackage{makecell}
\usepackage{mathtools}
\usepackage{float}
\usepackage{gensymb}
\usepackage{booktabs}
\usepackage{bm}
\usepackage{algorithm}
\usepackage{algpseudocode}
\usepackage{varwidth}
\usepackage[export]{adjustbox}
\def\BibTeX{{\rm B\kern-.05em{\sc i\kern-.025em b}\kern-.08em
    T\kern-.1667em\lower.7ex\hbox{E}\kern-.125emX}}

\usepackage{hyperref}
\hypersetup{
    colorlinks=true,
    linkcolor=blue,
}

\begin{document}

\newcommand{\bpsi}{\boldsymbol{\Psi}}
\newcommand{\bphi}{\boldsymbol{\phi}}
\newtheorem{remark}{Remark}

\newcommand{\blue}[1]{\textcolor{blue}{#1}}

\bstctlcite{IEEEexample:BSTcontrol}

\title{Distributed MIMO Networks with Rotary ULAs for Indoor Scenarios under Rician Fading}

\author{Eduardo N. Tominaga, \textit{Student Member, IEEE}, Onel L. A. López, \textit{Senior Member, IEEE}, Tommy Svensson, \textit{Senior Member, IEEE}, Richard D. Souza, \textit{Senior Member, IEEE}, Hirley Alves, \textit{Member, IEEE}
\thanks{
This research was financially supported by Research Council of Finland (former Academy of Finland), 6Genesis Flagship (grant no. 346208), European Union’s Horizon 2020 research and innovation programme (EU-H2020), Hexa-X-II (grant no. 101095759) project, the Finnish Foundation for Technology Promotion, and in Brazil by CNPq (305021/2021-4, 402378/2021-0) and RNP/MCTIC 6G Mobile Communications Systems (01245.010604/2020-14). (\textit{Corresponding author: Eduardo N. Tominaga})

Eduardo N. Tominaga, Onel L. A. López, and Hirley Alves are with the Centre for Wireless Communications (CWC), University of Oulu, Finland. (E-mail: \{eduardo.noborotominaga,onel.alcarazlopez,hirley.alves\}@oulu.fi).

Tommy Svensson is with the Department of Electrical Engineering, Chalmers University of Technology, 412 96 Gothenburg, Sweden (E-mail: tommy.svensson@chalmers.se).

Richard Demo Souza is with the Department of Electrical and Electronics Engineering, Federal University of Santa Catarina (UFSC), Florian\'{o}polis, 88040-370, Brazil. (E-mail: richard.demo@ufsc.br).
}
}

\maketitle

\begin{abstract}
The Fifth-Generation (5G) wireless communications networks introduced native support for Machine-Type Communications (MTC) use cases.
Nevertheless, current 5G networks cannot fully meet the very stringent requirements regarding latency, reliability, and number of connected devices of most MTC use cases. Industry and academia have been working on the evolution from 5G to Sixth Generation (6G) networks. One of the main novelties is adopting Distributed Multiple-Input Multiple-Output (D-MIMO) networks. However, most works studying D-MIMO consider antenna arrays with no movement capabilities, even though some recent works have shown that this could bring substantial performance improvements. In this work, we propose the utilization of Access Points (APs) equipped with Rotary Uniform Linear Arrays (RULAs) for this purpose. Considering a spatially correlated Rician fading model, the optimal angular position of the RULAs is jointly computed by the central processing unit using particle swarm optimization as a function of the location of the active devices. Considering the impact of imperfect location estimates, our numerical results show that the RULAs's optimal rotation brings substantial performance gains in terms of mean per-user spectral efficiency. The improvement grows with the strength of the line-of-sight components of the channel vectors. Given the total number of antenna elements, we study the trade-off between the number of APs and the number of antenna elements per AP, revealing an optimal number of APs for the cases of APs equipped with static ULAs and RULAs.
\end{abstract}


\section{Introduction}

\par The Fifth Generation (5G) of wireless communication networks is currently under deployment and in advanced phases of standardization worldwide. Besides offering enhanced coverage capabilities and data rates for human-type communication applications such as mobile broadband internet connectivity, which corresponds to a natural evolution of previous generations, 5G also offers services for a variety of Machine-Type Communication (MTC) applications \cite{shariatmadari2015}. 5G MTC use cases are traditionally split into two broad categories: massive MTC (mMTC) and Ultra-Reliable Low Latency Communications (URLLC), also known as critical MTC (cMTC). The former aims to provide wireless connectivity to a massive number of low-power and low-complexity devices in applications with moderate data rates, latency, and reliability requirements. The latter category comprises applications with very stringent requirements in terms of latency and reliability. For instance, in an industrial setting, mMTC would correspond to sensors and actuators monitoring and controlling non-critical processes. In contrast, cMTC connectivity would enable critical control and safety systems to operate wirelessly \cite{wen2022}.

\par The requirements of the different mMTC and cMTC use cases are foreseen to become even more stringent in future Sixth Generation (6G) networks. Thus, academia and industry have joined efforts in enhancing current technologies and developing new ones \cite{uusitalo2021}. One of the novelties that have mostly attracted the attention of academia and industry is Distributed Multiple-Input Multiple Output (D-MIMO), often referred to as Cell-Free massive MIMO \cite{ngo2017,chen2018,interdonato2019}. Instead of having a single base station equipped with several antenna elements serving a coverage area, the antenna elements are distributed among multiple Access Points (APs) over the coverage area. The APs are connected to a common Central Processing Unit (CPU) through fronthaul links. Such an approach provides a more uniform wireless coverage, enhancing network performance.

\par The vast majority of works investigating the performance of D-MIMO networks consider fixed antenna arrays, i.e., antenna arrays with no movement capabilities. Nevertheless, the idea of antenna arrays that can move has gained attention among the research community \cite{lozano2021,onel_francisco_2021,onel2021,lin2023,zubow2023,zhu2023,xiao2023}. According to recent works \cite{zhu2023,zhu2024,xiao2023}, moving and/or rotating antennas can take advantage of spatially varying channel conditions within a confined area. Indeed, by altering the position and/or orientation of the antenna, better channel conditions can be achieved. In MTC scenarios, devices are typically deployed in fixed locations, while the surrounding environment may change over time, leading to slowly varying wireless channels. In use cases without mobility, narrowband MTC offers limited time and frequency diversity, which restricts the potential for improving data rates and transmission reliability. In such situations, antennas with movement capabilities are a promising solution for achieving higher spatial diversity gains.
The related works will be discussed in more detail in the following subsection.

\subsection{Related Works}

\par The utilization of antenna arrays with movement capabilities is not new. For instance, the authors in \cite{li2018} proposed a Direction of Arrival (DOA) estimation method that utilizes a Rotary Uniform Linear Array (RULA) and achieves satisfactory performance for under-determined DOA estimations, where the number of source signals can be larger than the number of receive antenna elements. The achievable rate performance of point-to-point Line-of-Sight (LoS) links with both the transmitter and receiver equipped with a RULA was studied in \cite{lozano2021} and shown to approach the LoS capacity at any desired Signal-to-Noise Ratio (SNR). López \textit{et. al.} \cite{onel_francisco_2021,onel2021} and Lin \textit{et. al.} \cite{lin2023} proposed the utilization of RULAs for wireless energy transfer. They studied a system where a power beacon equipped with a RULA constantly rotates and transmits energy signals in the downlink to several low-power devices. The devices harvest energy from the transmitted signal to recharge their batteries. The authors in \cite{zubow2023} developed and tested a prototype for hybrid mechanical-electrical beamforming for mmWave WiFi. Their experimental results in a point-to-point setup showed that the optimal rotation of the antenna array can bring significant throughput improvements for both LoS and Non-LoS (NLoS) scenarios.

\par More recently, movable antennas, which can move along one or several directions within a confined area, have been proposed \cite{zhu2023,xiao2023}. The main drawback of their utilization is that each movable antenna requires at least two cables and two servo motors, representing high deployment, operation, and maintenance costs. A different case is that of UAVs operating as flying base stations \cite{amponis2022}, which can also be interpreted as APs with movement capabilities. Their most notable advantage is that they present several degrees of freedom for movement since a UAV can be positioned at any point of the coverage area, at any height, and their position can be easily changed. However, their main drawbacks are limited carrying capacity, very high power consumption, and the consequent need for frequent recharges \cite{mohsan2023}.

\par To the best of the authors' knowledge, this paper is the first work that investigates a D-MIMO network implemented with APs equipped with RULAs\footnote{Preliminary results of this work were published in the conference version \cite{tominaga2024}. In that work, we consider a similar scenario but only a single AP. Since the optimization problem studied in that work has only one variable (we optimize the angular position of only a single AP), brute force search was used instead of PSO.}. The most important advantages of the RULAs, when compared to the alternative approaches, are the lower deployment, operation, and maintenance costs since each AP requires a single servo-motor to rotate its Uniform Linear Array (ULA).

\begin{table}[]
    \centering
    \caption{List of Acronyms}
    \begin{tabular}{l l}
        \toprule
        \textbf{Acronym} & \textbf{Definition} \\
        \midrule
        AP & Access Point\\
        CPU & Central Processing Unit\\
        CSI & Channel State Information\\
        D-MIMO & Distributed MIMO\\
        LoS & Line-of-Sight\\
        MIMO & Multiple-Input Multiple-Output\\
        MTD & Machine-Type Device\\
        NLoS & Non-LoS\\
        PSO & Particle Swarm Optimization\\
        RA & Random Access\\
        RULA & Rotary ULA\\
        SE & Spectral Efficiency\\
        SNR & Signal-to-Noise Ratio\\
        SINR & Signal-to-Interference-plus-Noise Ratio\\
        ULA & Uniform Linear Array\\
        ZF & Zero Forcing\\
        \bottomrule
    \end{tabular}    
    \label{tableAcronyms}
\end{table}

\subsection{Contributions and Organization of the Paper}

\par In this work, we evaluate the performance of D-MIMO networks for indoor industrial networks. We consider the uplink of a fully centralized system, where all APs simultaneously serve all the active Machine-Type Devices (MTDs). All APs are equipped with static ULAs or RULAs and connected to a common CPU through fronthaul links. The contributions of this work are listed below:
\begin{itemize}
    \item We propose a scheme for joint configuration of the angular positions of all the RULAs in the system to maximize the mean per-user achievable Spectral Efficiency (SE). We assume that the network obtains estimates of the locations of the active devices (i.e., the devices that seek to transmit data in the uplink) before the data transmission phase and using some indoor localization method. The estimates are utilized to compute the optimal angular positions of the RULAs of the APs using the Particle Swarm Optimization (PSO).
    \item Adopting a spatially correlated Rician fading channel model, we evaluate the performance of the different setups considering different values of the Rician factor. Our numerical results show that the optimal rotation of the RULAs can substantially improve the mean-per-user achievable SE and that the performance gains grow with the strength of the LoS components of the channel vectors.
    \item We propose a localization error model and define different levels of localization accuracy based on the requirements specified by the Third Generation Partnership Project (3GPP) standards. The system's performance is evaluated as a function of the localization accuracy. Our numerical results show that the optimal rotation of the RULAs brings performance gains even when the system's localization accuracy is poor.
    \item We evaluate the trade-off between the number of APs and antennas per AP given a total number of antenna elements. Our numerical results show that for both the cases of APs equipped with static ULAs and APs equipped with RULAs, there is an optimal number of APs that achieves the highest SE values.
\end{itemize}

\par Table \ref{tableAcronyms} lists the acronyms used throughout this paper alphabetically. This paper is organized as follows. Section \ref{systemModel} presents the system model. Section \ref{localizationErrorModel} introduces the localization error model. Section \ref{optimalAngularPositions} describes the proposed mechanism for optimizing the angular position of the RULAs. Section \ref{numericalResults} presents and discusses the numerical results. Finally, Section \ref{conclusions} concludes the paper.

\par \textbf{Notation:} lowercase boldface letters denote column vectors, while boldface uppercase letters denote matrices. $a_i$ is the $i$-th element of the column vector $\textbf{a}$, while $\textbf{a}_i$ is the $i$-th column of the matrix $\textbf{A}$.
$\textbf{I}_M$ is the identity matrix with size $M\times M$. The superscripts $(\cdot)^T$ and $(\cdot)^H$ denote the transpose and the conjugate transpose of a vector or matrix, respectively. The scalar quantity's magnitude or the set's cardinality is denoted by $|\cdot|$, and $ \Vert\cdot\rVert$ denotes the Euclidian norm of a vector (2-norm). We denote the one-dimensional uniform distribution with bounds $a$ and $b$ by $\mathcal{U}(a,b)$. We denote the multivariate Gaussian distribution with mean $\mathbf{a}$ and covariance $\mathbf{B}$ by $\mathcal{N}(\mathbf{a},\mathbf{B})$.

\section{System Model}
\label{systemModel}

\par
\begin{figure}[t]
    \centering
    \includegraphics[scale=0.5]{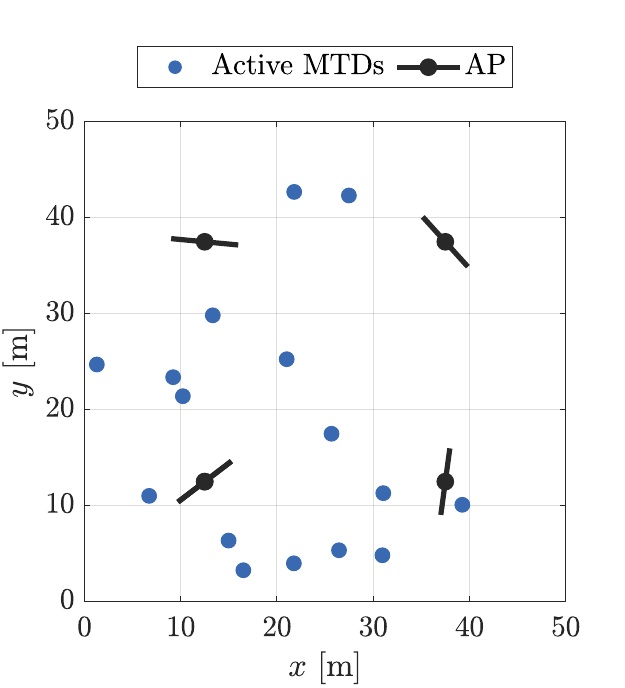}
    \caption{Illustration of the considered system model for $l$~$=$~$50$~m, $Q=4$, and $K=16$.}
    \label{illustrationSystemModel}
\end{figure}

 Similar to simulation setups of many relevant and related works (see, for instance, \cite{ngo2017,nayebi2017,femenias2020,chen2021}),  we consider a square coverage area with dimensions $l\times l\;\text{m}^2$. $Q$ APs serve the coverage area, each with a RULA of $S$ half-wavelength spaced antenna elements. The APs are at height $h_{\text{AP}}$. The $Q$ APs jointly serve $K$ MTDs that are active simultaneously and transmit data in the uplink. The locations of the MTDs are fixed, that is, we consider a scenario with no mobility. Let $(x_k,y_k)$ denote the coordinates of the $k$-th MTD. For simplicity, we consider that the antenna elements of all MTDs are located at the same height $h_{\text{MTD}}$ \cite{ngo2017,chen2018}. The system model is illustrated in Fig.~\ref{illustrationSystemModel}.

\par We consider fully centralized\footnote{In this work, for the sake of simplicity and to highlight the novelty of our proposed framework, we assume that all APs are perfectly calibrated and synchronized, a common assumption in several relevant works (e.g., \cite{nayebi2017,ngo2017,femenias2020,chen2021}). Note that imperfect synchronization can affect any D-MIMO system in both cases of APs equipped with static or rotary/movable antennas. Nevertheless, there is no indication that any alternatives would be more or less sensitive to synchronization.} processing: all the APs are connected to a common CPU through high-capacity fronthaul connections. The CPU is responsible for performing the following signal-processing tasks:
\begin{enumerate}
    \item MTD's activity detection and identification.
    \item Computation of the estimates of the locations of the active MTDs using an indoor localization method.
    \item Computation of the optimal angular positions of the RULAs based on the estimated locations of the active MTDs.
    \item Scheduling of the uplink data transmissions.
    \item Computation of the Channel State Information (CSI)  of the links between any AP and any MTD using pilot sequences.
    \item Computation of centralized receive combining vectors based on the global CSI knowledge.
    \item Linear centralized uplink data decoding.
\end{enumerate}
Further details on each of these tasks are provided in the subsequent sections of this work.

\subsection{Channel Model}

\par We adopt a spatially correlated Rician fading channel model \cite{ozdogan2019}. Let $\textbf{h}_{kq}\in\mathbb{C}^{S\times1}$ denote channel vector between the $k$-th MTD and the $q$-th AP, which can be modeled as \cite{dileep2021}
\begin{equation}
    \label{channelVector}
    \textbf{h}_{kq}=\sqrt{\dfrac{\kappa}{1+\kappa}}\textbf{h}_{kq}^{\text{los}} + \sqrt{\dfrac{1}{1+\kappa}}\textbf{h}_{kq}^{\text{nlos}},
\end{equation}
where $\kappa$ is the Rician factor, $\textbf{h}_{kq}^{\text{los}}\in\mathbb{C}^{S\times1}$ is the deterministic LoS component, and $\textbf{h}_{kq}^{\text{nlos}}\in\mathbb{C}^{S\times1}$ is the random NLoS component.

\par The deterministic LoS component is given by
\begin{equation}
    \label{losComponent}
    \textbf{h}_{kq}^{\text{los}}=\sqrt{\beta_{kq}}
    \begin{bmatrix}
        1\\
        \exp(-j2\pi\Delta\sin(\phi_{kq}))\\
        \exp(-j4\pi\Delta\sin(\phi_{kq}))\\
        \vdots\\
        \exp(-j2\pi(S-1)\Delta\sin(\phi_{kq}))\\
    \end{bmatrix},
\end{equation}
where $\beta_{kq}$ is the power attenuation owing to the distance between the $k$-th MTD and the $q$-th AP, and $\phi_{kq}\in[0,2\pi]$ is the azimuth angle relative to the boresight of the ULA of the $q$-th AP. Meanwhile, the random NLoS component is distributed as
\begin{equation}
    \textbf{h}_{kq}^{\text{nlos}}\sim\mathcal{CN}(\textbf{0},\textbf{R}_{kq}).
\end{equation}
Note that
\begin{equation}
    \textbf{h}_{kq}\sim\mathcal{CN}\left(\sqrt{\dfrac{\kappa}{1+\kappa}}\textbf{h}_{kq}^{\text{los}},\dfrac{\textbf{R}_{kq}}{\kappa+1}\right),
\end{equation}
where $\textbf{R}_{kq}\in\mathbb{C}^{M\times M}$ is the positive semi-definite covariance matrix describing the spatial correlation of the NLoS components.

\par The spatial covariance matrices can be (approximately) modeled using the Gaussian local scattering model \cite[Sec. 2.6]{bjornson2017}. Specifically, the $s$-th row, $m$-th column element of the correlation matrix is
\begin{equation}
\begin{split}
    [\textbf{R}_{kq}]_{s,m}=\dfrac{\beta_{kq}}{N}\sum_{n=1}^N\exp[j\pi(s-m)\sin(\psi_{kq,n})] \\
    \times \exp\left\{-\dfrac{\sigma_\phi^2}{2}[\pi(s-m)\cos(\psi_{kq,n})]^2 \right\},
\end{split}
\end{equation}
where 
$N$ is the number of scattering clusters, $\psi_{kq,n}$ is the nominal angle of arrival for the $n$-th cluster, and $\sigma_\psi$ is the angular standard deviation.

\subsection{Signal Model}

\par Herein, we consider an uplink scenario where $K$ active MTDs simultaneously transmit signals to the $Q$ APs. Let $M=QS$ denote the total number of receive antenna elements. The collective vector of wireless channel coefficients between the $k$-th MTD and the $Q$ APs is $\textbf{h}_k=[\textbf{h}_{k1}^T,\textbf{h}_{k2}^T,\ldots,\textbf{h}_{kQ}^T]^T \in \mathbb{C}^{M\times1}$.

\par The matrix $\textbf{H}\in\mathbb{C}^{M\times K}$ containing the channel vectors of the $K$ MTDs can be written as
\begin{equation}
    \textbf{H}=[\textbf{h}_1,\textbf{h}_2,\ldots,\textbf{h}_K].
\end{equation}
The $M\times 1$ collective received signal vector can be written as
\begin{equation}
    \textbf{y}=\sqrt{p}\textbf{H}\textbf{x}+\textbf{n},
\end{equation}
where $p$ is the fixed uplink transmit power that is the same for all MTDs, $\textbf{x}\in\mathbb{C}^{K\times 1}$ is the vector of symbols simultaneously transmitted by the $K$ MTDs, and $\textbf{n}\in\mathbb{C}^{M\times 1}$ is the vector of additive white Gaussian noise samples such that $\textbf{n}\sim\mathcal{CN}(\textbf{0}_{M\times1},\sigma^2_n\textbf{I}_M)$.

\par Let $\textbf{V}\in\mathbb{C}^{M\times K}$ be a linear detector matrix used for the joint decoding of the signals transmitted from the $K$ MTDs at all the APs. The received signal after the linear detection operation is split in $K$ streams and given by
\begin{equation}
    \textbf{r}=\textbf{V}^H\textbf{y}=\sqrt{p}\textbf{V}^H\textbf{H}\textbf{x}+\textbf{V}^H\textbf{n}.
\end{equation}
Let $r_k$ and $x_k$ denote the $k$-th elements of $\textbf{r}$ and $\textbf{x}$, respectively. Then, the received signal corresponding to the $k$-th MTD is
\begin{equation}
    \label{r_k}
    r_k=\underbrace{\sqrt{p}\textbf{v}_k^H\textbf{h}_kx_k}_\text{Desired signal} + \underbrace{\sqrt{p}\textbf{v}_k^H\sum_{k'\neq k}^K \textbf{h}_{k'}x_{k'}}_\text{Inter-user interference} + \underbrace{\textbf{v}_k^H\textbf{n}}_\text{Noise},
\end{equation}
where $\textbf{v}_k$ and $\textbf{h}_k$ are the $k$-th columns of the matrices $\textbf{V}$ and $\textbf{H}$, respectively.

\par From (\ref{r_k}), the Signal-to-Interference-plus-Noise Ratio (SINR) of the uplink transmission from the $k$-th MTD to all the APs is given by
\begin{equation}
    \label{gamma_k}
    \gamma_k=\dfrac{p|\textbf{v}_k^H\textbf{h}_k|^2}{p\sum_{k'\neq k}^K |\textbf{v}_k^H\textbf{h}_{k'}|^2+\sigma^2_n\lVert\textbf{v}_k^H\rVert^2}.
\end{equation}
The receive combining matrix $\textbf{V}$ is computed as a function of the matrix of estimated channel vectors $\hat{\textbf{H}}\in\mathbb{C}^{M\times K}$, $\hat{\textbf{H}}=[\hat{\textbf{h}}_1,\ldots,\hat{\textbf{h}}_K]$. In this work, we adopt the centralized Zero Forcing (ZF) combining scheme\footnote{Centralized ZF for distributed massive MIMO in indoor industrial scenarios was also adopted in \cite{alonzo2021}. Note that the performance obtained with ZF approaches that with MMSE in the high SINR regime \cite{paulraj2003}, with the advantage of not requiring statistical knowledge of the noise variance or interference.}. The receive combining matrix is computed as \cite{liu2020}
\begin{equation}
    \label{precodingMatrices}
    \textbf{V}=\hat{\textbf{H}}(\hat{\textbf{H}}^H\hat{\textbf{H}})^{-1}.
\end{equation}

\subsection{Performance Metrics}
\label{performanceMetrics}

\par To illustrate the improvement in the deterministic LoS links owing to the optimal rotations of the RULAs, we adopt the mean per-user achievable SE as the performance metric. The achievable uplink SE of the $k$-th MTD is
\begin{equation}
    \label{per-user-achievable-SE}
    R_k=\mathbb{E}_{\textbf{H}}\{\log_2(1+\gamma_k)\}.
\end{equation}
Then, the mean per-user achievable uplink SE is obtained by averaging over the achievable uplink SE of all the $K$ active MTDs, i.e.,
\begin{equation}
    \label{mean-per-user-achievable-SE}
    \Bar{R}=\dfrac{1}{K}\sum_{k=1}^K R_k.
\end{equation}

\subsection{Imperfect CSI Model}

\par Assuming $\tau_p$ orthogonal pilot sequences during the uplink data transmission~phase, such that $\tau_p\geq K$, and least squares channel estimation, the true channel realizations and the channel estimation errors are uncorrelated \cite{onel2022}. Then, the estimated channel vector of the $k$-th MTD, $\Hat{\textbf{h}}_k\in\mathbb{C}^{M\times1}$, can be modeled as the sum of the true channel vector plus a random error vector as \cite{wang2012,eraslan2013,onel2022}
\begin{equation}
    \label{estimatedChannelVectors}
    \hat{\textbf{h}}_k=\textbf{h}_k+\Tilde{\textbf{h}}_k,
\end{equation}
where $\Tilde{\textbf{h}}_k\sim\mathcal{CN}(\textbf{0},\sigma_{\text{csi}}^2\textbf{I}_M)$ is the vector of channel estimation errors.

\par The parameter $\sigma_{\text{csi}}^2$ indicates the quality of the channel estimates. Let $\rho=p/\sigma^2_n$ denote the transmit SNR. We also assume that the duration of the uplink pilot transmission phase is equal to $\tau_p$ symbols. Then, the variance of the channel estimation errors can be modeled as a decreasing function of $\rho$ as \cite{wang2012,eraslan2013,onel2022}
\begin{equation}
    \sigma_{\text{csi}}^2=\dfrac{1}{\tau_p\rho}.
\end{equation}
Note that the channel estimation error depends only on the uplink transmit power, average receive noise power and number of orthogonal pilots, thus it is the same for all devices. The consequence of imperfect CSI is a slight decrease in the mean per-user achievable SEs compared to the case of perfect CSI for both the cases of static ULAs and RULAs.

\begin{figure*}[t]
    \centering
    \includegraphics[scale=0.6]{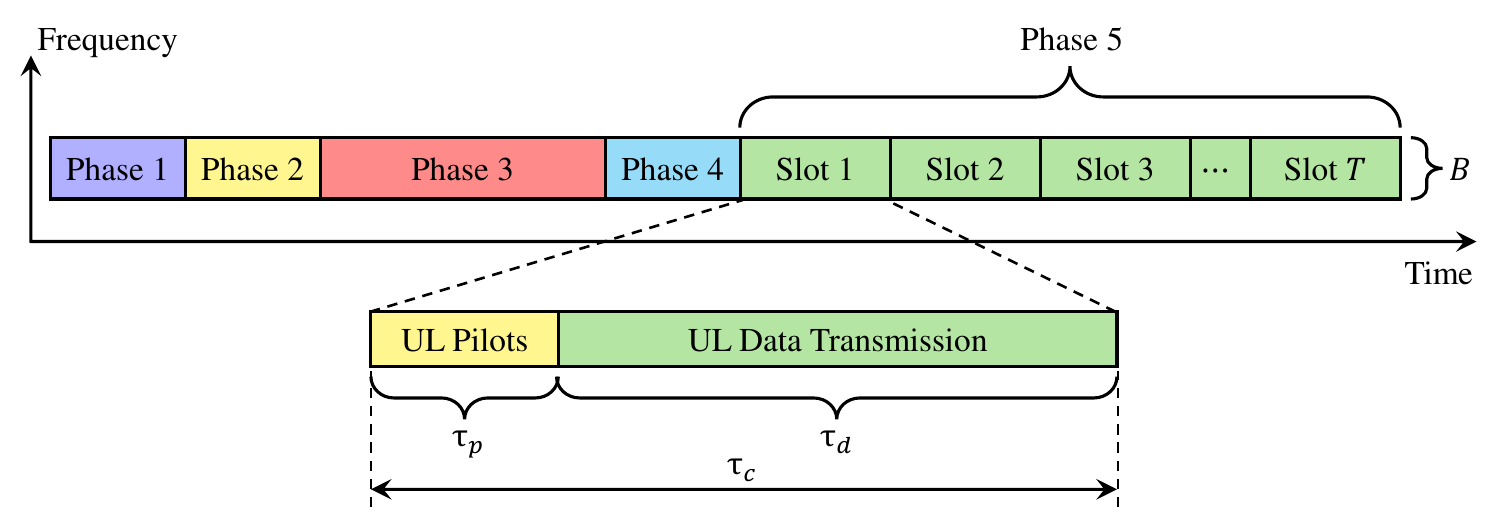}
    \caption{Illustration of the proposed framework for the optimization of the angular positions of the RULAs and uplink data transmission. Note that the phases have different duration.}
    \label{illustrationProtocol}
\end{figure*}

\subsection{Proposed Framework}
\label{threePhaseRandomAccess}

\par In this subsection, we describe our proposed framework for the optimization of the rotation of the $Q$ APs and subsequent uplink data transmission. We assume that the same subset of devices is active during a period of time that spans multiple consecutive time slots. For instance, the devices could be video surveillance cameras \cite{dawy2017} or smart utility meters \cite{nardelli2016}. Therefore, we assume that the optimization of the angular positions of the RULAs is performed before the data transmission phase, thus the angular positions of the RULAs is kept constant during multiple consecutive time slots. Inspired by the three-phase scheduled random access scheme from \cite{kang2022}, our framework has the following phases:
\begin{enumerate}
    \item Active MTDs, i.e., MTDs seeking to send data to the network, transmit non-orthogonal uplink pilots for activity detection.
    \item The network identifies the set of active MTDs and, utilizing some indoor localization techniques \cite{liu2007,yassin2017,zafari2019}, obtains estimates of the locations of the active MTDs.
    \item The network assumes pure LoS propagation and utilizes the estimated locations of the MTDs and location-based beamforming to compute its optimal rotation and/or position. The RULAs have their angular positions updated accordingly.
    \item The network broadcasts a common downlink feedback message to assign each device an orthogonal pilot sequence.
    \item The MTDs transmit their orthogonal pilot sequences and data during multiple time slots. The uplink orthogonal pilots are used to compute the CSI estimates shown in (\ref{estimatedChannelVectors}) for each coherence time interval, which are then used to compute the ZF receive combining vectors in (\ref{precodingMatrices}). 
\end{enumerate}

\par The proposed framework is illustrated in Fig. \ref{illustrationProtocol}. Note that phase 5 extends over $T$ time slots. Within each time slot, the $K$ active devices transmit simultaneously. Each time slot is as long as the channel's coherence time interval and contains $\tau_c$ symbols. The first $\tau_p$ symbols of each time slot are used for the transmission of the orthogonal pilot sequences, while the remaining $\tau_d=\tau_c-\tau_p$ symbols are utilized for uplink data transmission. The numerical results presented in this work, that is, the mean per-user achievable SE presented in Section II-\ref{performanceMetrics}, correspond to the performance achieved during phase~5.

\section{Localization Error Model}
\label{localizationErrorModel}

\par As mentioned in Section \ref{threePhaseRandomAccess}, the CPU utilizes the information about the locations of the MTDs to compute the optimal rotations of the RULAs. However, the estimates of the locations are not perfect in practical systems. In this section, we propose a mathematical model for the localization error and relate it to the localization accuracy requirements proposed by 3GPP.

\par Considering that all the MTDs are located at the same height $h_{\text{MTD}}$, the imperfect localization impairment refers to the uncertainty on the location of the MTDs only on the horizontal plane. Let $\textbf{p}_k=(x_k,y_k)$ denote the true location of the $k$-th MTD, and $\hat{\textbf{p}}_k=(\hat{x}_k,\hat{y}_k)$ denote the estimated location. The localization error vector associated to the $k$-th MTD becomes
\begin{equation}
   \textbf{e}_k=\textbf{p}_k-\hat{\textbf{p}}_k=(x_{e,k},y_{e,k}),
\end{equation}
where $x_{e,k}=x_k-\hat{x}_k$ and $y_{e,k}=y_k-\hat{y}_k$ are the $x$ and $y$ components of the localization error vector, respectively. We can rewrite the localization error vector in polar form as
\begin{equation}
   \textbf{e}_k=r_{e,k} \angle \theta_{e,k},
\end{equation}
where $r_{e,k}=\sqrt{x_{e,k}^2+y_{e,k}^2}$ and $\theta_{e,k}=\tan^{-1}(y_{e,k}/x_{e,k})$ are the magnitude and phase of the localization error vector, respectively.

\par Aiming at generality (that is, to not introduce any additional assumption or bias related to indoor localization), we model the localization error as a bivariate Gaussian distribution since it is the least informative\footnote{In information theory and statistics, ``least informative'' refers to the distribution with the maximum entropy (i.e., uncertainty) for a given variance.} distribution for any given variance \cite{gustafsson2005}. Specifically, we assume the localization error follows a bivariate Gaussian distribution with mean $\bm{\mu}=[\textbf{0}\;\textbf{0}]^T$ and covariance matrix $\bm{\Sigma}=\sigma_e^2\textbf{I}_2$ as in \cite{zhu2022}. Thus, $x$ and $y$ components of the localization error vector follow a Normal distribution:
\begin{equation}
    x_{e,k},y_{e,k}\sim\mathcal{N}(0,\sigma_e^2).
\end{equation}
Moreover, the magnitude of the localization error vector follows a Rayleigh distribution:
\begin{equation}
    f_R(r|\sigma_e^2)=\dfrac{r}{\sigma_e^2}\exp\left(-\dfrac{r^2}{2\sigma_e^2}\right),\;r\geq 0,
\end{equation}
where $\sigma_e^2$ is the scaling parameter. Finally, the angle of the localization error vector follows a uniform distribution:
\begin{equation}
    \theta_{e,k}\sim\mathcal{U}(-\pi,\pi).
\end{equation}

\par Let $r_{e,k}$, i.e., the radius of the localization error vector, represent the localization accuracy. Considering the bivariate Gaussian model, the localization accuracy follows the Rayleigh distribution. Thus, we have
\begin{align}
    & \mathbb{E}\{r_{e,k}\}=\sigma_e\sqrt{\pi/2},\\
    & \text{Var}\{r_{e,k}\}=\dfrac{4-\pi}{2}\sigma_e^2.
\end{align}

\par In 3GPP standards, the localization accuracy requirements are typically specified in terms of the 95\%-quantile \cite{3gpp2019}. The $F$-quantile of a Rayleigh distribution with scaling parameter $\sigma_e$ is given by
\begin{equation}
    Q(F;\sigma_e)=\sigma_e\sqrt{-2\ln(1-F)}.
\end{equation}
Considering the proposed localization error model and based on the 3GPP specifications, we list examples of localization accuracy requirements for pessimistic, reasonable, and optimistic scenarios and their corresponding values of $\sigma_e^2$ in Table \ref{localizationRequirements}. In 3GPP Rel-16, the localization accuracy requirement specified for indoor scenarios was $3$ m \cite{3GPP-rel16}. In 3GPP Rel-17, this requirement was further reduced to $20$ cm \cite{3GPP-rel17}. The localization accuracy requirement for beyond-5G and 6G networks in indoor factory scenarios is expected to be in the order of a few centimeters \cite{nikonowicz2024}.

\begin{table}[t]
    \centering
    \caption{Examples of localization accuracy requirements for pessimistic, optimistic, and reasonable scenarios \cite{3GPP-rel16,3GPP-rel17,nikonowicz2024}.}
    \begin{tabular}{l l l}
        \toprule
        $\phantom{-}\sigma_e^2$ & 95\% Quantile & Scenario \\
        \midrule
        $\phantom{-}5$ dB & $4.5$ m & Pessimistic \\
        $\phantom{-}1.76$ dB & $3$ m & Reasonable \\
        $-7.8$ dB & $1$ m & Reasonable \\
        $-28$ dB & $10$ cm & Optimistic \\
        $-45$ dB & $1$ cm & Optimistic \\
        \bottomrule
    \end{tabular}    
    \label{localizationRequirements}  
\end{table}

\section{Optimal Angular Positions of the RULAs}
\label{optimalAngularPositions}

\par In each time slot, $K$ distinct MTDs are active. Our previous work \cite{tominaga2024} shows a distinct optimal angular position for the RULA for each subset of $K$ locations of active MTDs. Moreover, high-speed industrial servo motors can change their angular positions within a few milliseconds\footnote{In this work, we assume that the rotation of the RULAs is performed only once and before the uplink data transmission phase. Consequently, the time required to rotate the RULAs does not yield channel aging or any other impact on the uplink data transmission. Nevertheless, timing requirements are crucial for the operation of our proposed system, and we intend to study them more thoroughly in future extensions of this work.} with very high precision \cite{Constar,Yaskawa}. Thus, they can track slow-to-moderate network dynamics in practice. Moreover, assuming the ULA is implemented as a microstrip patch antenna array, similar to the prototypes studied in \cite{host2013,kothapudi2019,binliu2020,tetsuya2021}, it will be lightweight. Hence, it will not impose a significant payload burden on the servo motor.

\par
\begin{figure}[t]
    \centering
    \begin{subfigure}{.23\textwidth}
        \centering
        \includegraphics[scale=0.45]{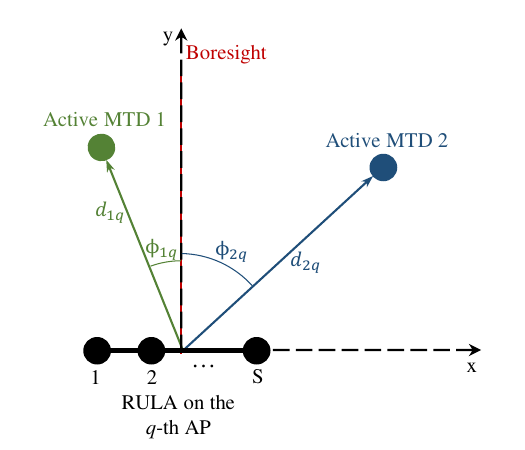}
        \caption{}
    \end{subfigure}%
    \begin{subfigure}{.23\textwidth}
        \centering
        \includegraphics[scale=0.45]{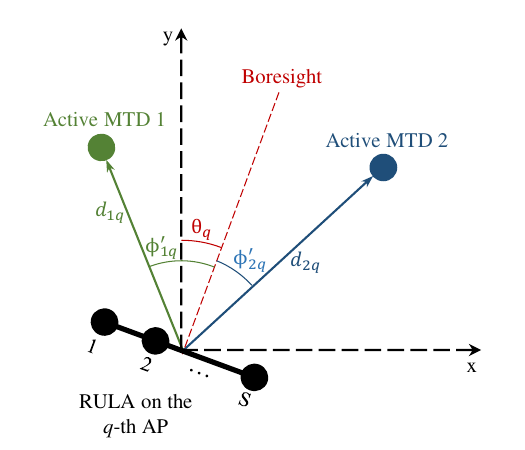}
        \caption{}
    \end{subfigure}
    \caption{Illustration of the RULA of the $q$-th AP and its angular positions concerning two active MTDs (a) before the rotation and (b) after the rotation.}
    \label{Illustration_RULA}
\end{figure}

 Let $\theta_q\in[-\pi,\pi]$ denote the rotation of the RULA of the $q$-th AP. A RULA and its angular position concerning two active MTDs, before and after its rotation, is illustrated in Fig. \ref{Illustration_RULA}. The angle between the $k$-th MTD and the $q$-th AP after the rotation of the RULA is given by \cite{onel_francisco_2021}
\begin{equation}
    \phi_{kq}':=\phi_{kq}+\theta_q.
\end{equation}

\par The optimal rotations of the RULAs $\theta_q\,;\forall q\in\{1,\ldots,Q\},$ are jointly computed by the CPU as a function of $\hat{\textbf{p}}_k,\;\forall k\in\{1,\ldots,K\}$. This computation can be written as the following optimization problem:
\begin{equation}
    \begin{aligned}
    & \text{maximize} && f(\theta_1,\ldots,\theta_Q\;|\;\Hat{\textbf{p}}_k,\;k\in\{1,\ldots,K\}) \\
    & \text{subject to} && 0\leq\theta_1,\ldots,\theta_Q\leq\pi,
    \end{aligned}
\end{equation}
where $f(\cdot)$ is the objective function to be maximized. Owing to the large number of variables to be jointly optimized and the high non-linearity of this problem, it is solved using particle swarm optimization (PSO), which will be presented in the next subsection.

\subsection{Particle Swarm Optimization}

\par PSO \cite[Ch. 16]{engelbrecht2007} is very suitable for solving problems where a function's global maximum or minimum is very difficult to find. The algorithm works over a population of candidate solutions, known as agents or particles, and moves them in the search space according to their current positions and velocities. Each particle's movement is influenced by its best-known position and the global best-known position in the search space. The local and global best positions are updated on each iteration, and the algorithm is expected to move the swarm of particles toward the optimal solution.

\par PSO was initially proposed in \cite{kennedy1995} and intended to simulate social behavior, such as the movement of organisms in a bird flock or fish school. It has been used to solve many optimization problems in communication systems, such as optimal deployment, node localization, clustering, and data aggregation in wireless sensor networks \cite{kulkarni2011}. It has also been used to design antennas with a specific desired side-lobe level or antenna element positions in a non-uniform array \cite{jin2007}. In communication systems, it has also been used to compute the optimal precoding vector that maximizes the throughput of a multi-user MIMO system \cite{shu2009}, to optimize the scheduling in the downlink of a multi-user MIMO system \cite{hei2009}, and to initialize the channel estimates for MIMO-OFDM receivers that jointly perform channel estimation and decoding \cite{knievel2011}.

\par The parameters of the PSO algorithm are listed in Table \ref{parametersPSO_1}, while its pseudo-code is in Algorithm \ref{PSO}. The inertial weight $w$ controls the particle's tendency to continue in its current direction, while $c_1$ and $c_2$ are the acceleration coefficients that control the influence of the personal and global best positions, respectively. 
In our optimization problem, the number of variables to be jointly optimized equals the number of APs, $Q$. Each particle is a candidate set of angular positions for the $Q$ RULAs, i.e. $\{\theta_1,\ldots,\theta_Q\}$, which corresponds to a point in the $Q$-dimensional space with boundaries $[0,\pi]^{Q\times1}$. The objective function depends on the estimates of the locations of the MTDs, i.e., $f(\Hat{\textbf{p}}_1,\ldots,\Hat{\textbf{p}}_K)$, and is described in the next Subsection. There are two possible termination criteria: i) the maximum number of iterations is reached, or ii) the relative change in the value of the global best over a predefined number of maximum stall iterations is less than the tolerance $\epsilon$.

\begin{table}[t]
    \centering
    \caption{Parameters of the PSO Algorithm}
    \begin{tabular}{l l}
        \toprule
        \textbf{Symbol} & \textbf{Parameter} \\
        \midrule
        $f(\cdot)$ & Objective function\\
        $x_i$ & Position of the $i$-th particle\\
        $v_i$ & Velocity of the $i$-th particle\\
        $w$ & Inertial weight\\
        $c_1$ & Cognitive constant\\
        $u_1,u_2$ & Random numbers between 0 and 1\\
        $c_2$ & Social constant\\
        $p_{b,i}$ & Personal best of the $i$-th particle\\
        $g_b$ & Global best \\
        \bottomrule
    \end{tabular}    
    \label{parametersPSO_1}
\end{table}

\par PSO is suitable for the obtaining the maximum global or minimum of functions since its particles evaluate the function at multiple points at the same. Thus, the optimization problem is solved in a parallelized fashion, which significantly reduces the required computation time. Besides, having multiple particles distributed on the domain of the function makes it harder for the problem to converge to a local minimum or maximum point. Nevertheless, the major drawbacks of PSO are \cite{parsopoulos2002,sedighizadeh2009,shami2022}:
\begin{enumerate}
    \item There is no guarantee that it will indeed converge to the global minimum or maximum point because of its stochastic nature. The particles may become clustered around a local optimum point early in the optimization process, leading to premature convergence;
    \item Depending on the problem, PSO may be very sensitive to the tuning of its parameters, such as the number of the particles, inertia weight, cognitive and social constants, and the maximum number of iterations;
    \item PSO may present a very high computational cost when dealing with high-dimensions optimization problems (i.e. problems with many optimization variables) and not perform well at the same time.
\end{enumerate}

\begin{algorithm}[t]
\caption{Particle Swarm Optimization}
\label{PSO}
\begin{algorithmic}[1]
    \State Initialize the swarm of particles uniformly distributed over the $Q$-th dimensional space $[0,\pi]^{Q\times1}$
    \State Initialize personal best positions as the current particle's positions: $p_{b,i}^0\leftarrow x_i^0$
    \While{the termination criterion is not met}
        \For{Each particle}
            \State Evaluate each particle's position considering the objective function
            \If{Current position is better than personal best}
                \State Update personal best position
            \EndIf
            \If{Current position is better than global best}
                \State Update global best position
            \EndIf
        \EndFor
        \For{each particle}
            \State \begin{varwidth}[t]{\linewidth} Update particles' velocities:\\
            $v_i^{t+1} \gets wv_i^t + c_1u_1^t (p_{b,i}^t - x_i^t) + c_2u_2^t (g_b^t - x_i^t)$ \end{varwidth}
            \State \begin{varwidth}[t]{\linewidth} Move the particles to their next positions:\\
            $x_i^{t+1} \gets x_i^t + v_i^t$ \end{varwidth}
        \EndFor
    \EndWhile
\end{algorithmic}
\end{algorithm}

\subsection{Location-Based Beamforming}

\par Before \textit{Phase 3} of the three-phase RA scheme, the CPU does not have any CSI information. Thus, the only information available for the computation of the optimal angular positions of the RULAs is the estimates of the locations of the MTDs that will be active in each time slot. Thus, we adopt a location-based beamforming \cite{maiberger2010,kela2016,yan2016,liu2016} approach to estimate the corresponding objective function value.

\par Herein, we assume that the locations of all the APs are perfectly known. Given $\Hat{\textbf{p}}_k,\;\forall k$, the CPU computes the estimates for the distances and azimuth angles between all the APs and the MTDs, i.e. $\hat{d}_{k,q}$ and $\hat{\phi}_{k,q}$, $\forall k,\;\forall q$. Then, it computes pseudo channel vectors assuming full-LoS propagation as
\begin{equation}
    \textbf{h}_{kq}^{\text{pseudo}}=\sqrt{\Hat{\beta}_{kq}}
    \begin{bmatrix}
        1\\
        \exp(-j2\pi\Delta\sin(\Hat{\phi}_{kq}))\\
        \exp(-j4\pi\Delta\sin(\Hat{\phi}_{kq}))\\
        \vdots\\
        \exp(-j2\pi(S-1)\Delta\sin(\Hat{\phi}_{kq}))\\
    \end{bmatrix},
\end{equation}
where $\Hat{\beta}_{kq}$ is the estimated large-scale fading coefficient, which is computed as a function of the estimated distance $\Hat{d}_{kq}$ considering a known channel model.

\par Then, receive combining vectors are computed as a function of the pseudo channel vectors according to (\ref{precodingMatrices}). Finally, the cost function is obtained by computing the mean per-user achievable SE utilizing the pseudo-channel vectors and the corresponding receive combining vectors in (\ref{gamma_k}), (\ref{per-user-achievable-SE}) and~(\ref{mean-per-user-achievable-SE}).

\begin{figure*}
    \centering
    \begin{subfigure}[b]{0.23\textwidth}
        \centering
        \includegraphics[scale=0.33]{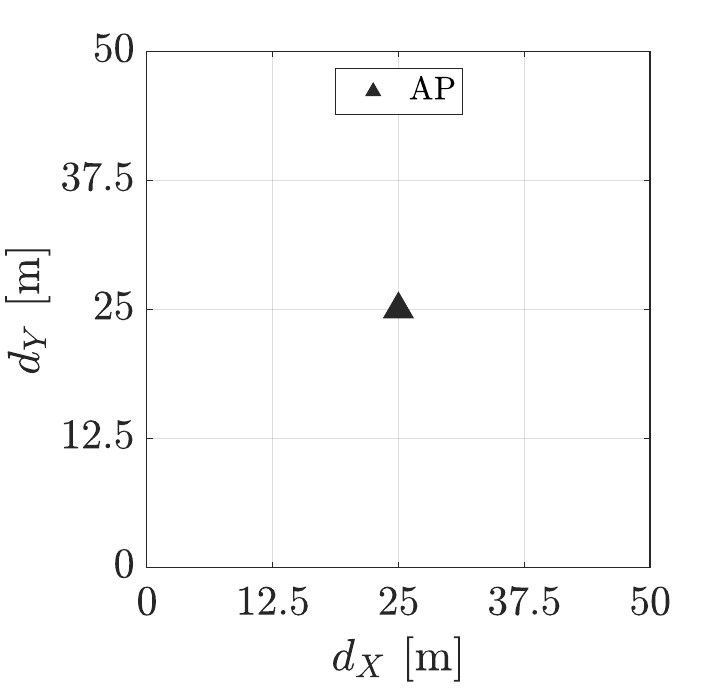}
        \caption{$Q=1$}    
    \end{subfigure}
    \hspace{0.1cm}
    \begin{subfigure}[b]{0.23\textwidth}  
        \centering 
        \includegraphics[scale=0.33]{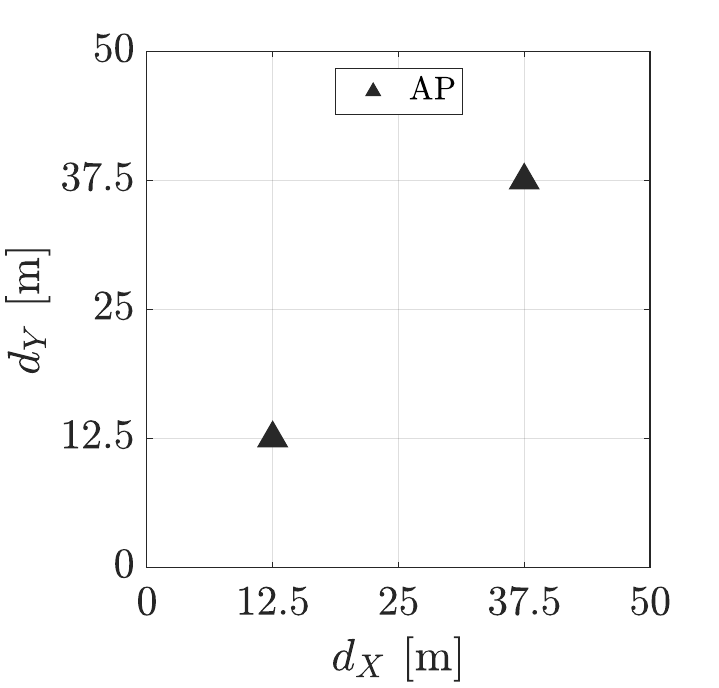}
        \caption{$Q=2$}  
    \end{subfigure}
    \begin{subfigure}[b]{0.23\textwidth}  
        \centering 
        \includegraphics[scale=0.33]{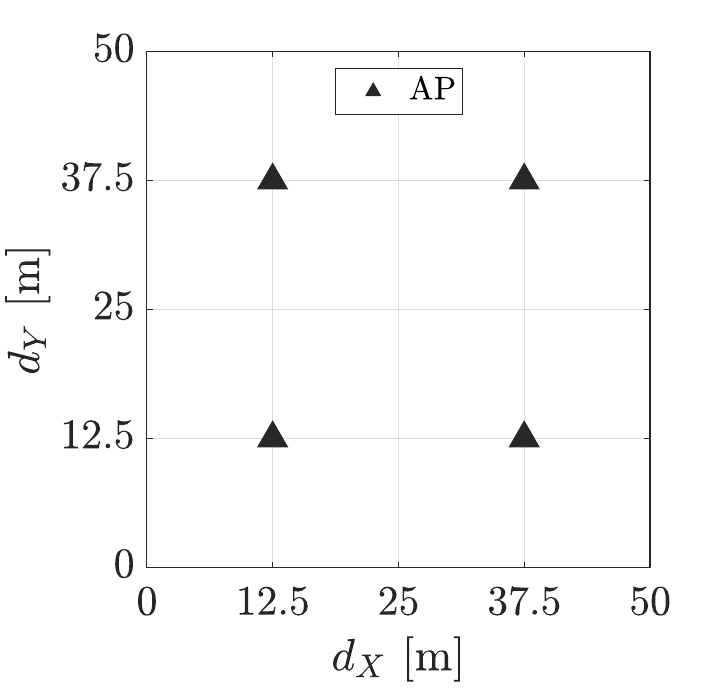}
        \caption{$Q=4$}  
    \end{subfigure}
    \begin{subfigure}[b]{0.23\textwidth}  
        \centering 
        \includegraphics[scale=0.33]{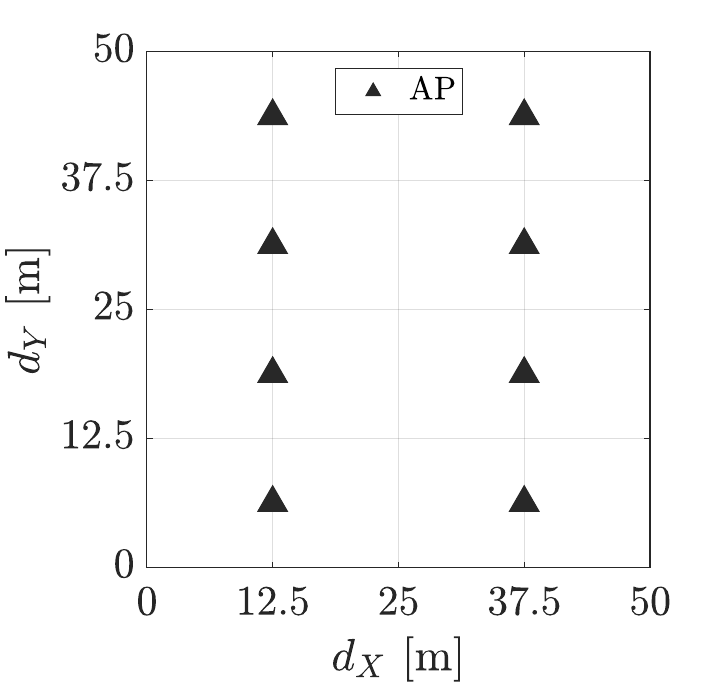}
        \caption{$Q=8$}  
    \end{subfigure}
    \caption{The square coverage area with $l=50$~m and the four different setups of APs considered in this work.}
    \label{setups}
\end{figure*}

\section{Numerical Results}
\label{numericalResults}

\par In this section, we present Monte Carlo simulation results that evince the performance improvements obtained by APs equipped with the proposed RULAs when compared to APs equipped with static ULAs. We keep the total number of antenna elements constant and study the trade-off between the number of APs and antenna elements per AP. The square coverage area\footnote{We opted to consider a square coverage area keeping in mind the shape of indoor industrial environments such as factory floors and warehouses. A square coverage area also facilitates the simulations, since it is simpler to uniformly distributed the APs this way. Moreover, we assuming circular or hexagonal areas should not lead to any relevant changes on the numerical results and conclusions. It should also be noted that in real-world scenarios, coverage areas can have irregular shapes.} with $l=50$~m and the four different setups of APs considered for the simulations are illustrated in Fig. \ref{setups}.

\subsection{Simulation Parameters}

\par The power attenuation due to the distance (in dB) is modelled using the log-distance path loss model as
\begin{equation}
    \beta_{kq}=-L_0-10\eta\log_{10}\left(\dfrac{d_{kq}}{d_0}\right),
\end{equation}
where $d_0$ is the reference distance in meters, $L_0$ is the attenuation owing to the distance at the reference distance (in dB), $\eta$ is the path loss exponent and $d_{kq}$ is the distance between the $k$-th device and the $q$-th AP in meters.

\par The attenuation at the reference distance is calculated using the Friis free-space path loss model and given by
\begin{equation}
    L_0=20\log_{10}\left(\dfrac{4\pi d_0}{\lambda}\right),
\end{equation}
where $\lambda=c/f_c$ is the wavelength in meters, $c$ is the speed of light and $f_c$ is the carrier frequency.

\par Unless stated otherwise, the values of the simulation parameters adopted in this work are listed in Table \ref{tableParameters}. Considering the typical values of $M$ and $h_{\text{AP}}$ defined in Table \ref{tableParameters}, every pair of AP and MTD are in the far-field (please refer to Appendix \ref{farFieldPropagation}). The active MTDs are uniformly distributed on the square coverage area, that is, $x_k,y_k\sim\mathcal{U}(0,l)$. Moreover, the adopted parameters for the PSO algorithm are listed in Table \ref{parametersPSO}.

\par The noise power (in Watts) is given by $\sigma^2_n=N_0BN_F$, where $N_0$ is the Power Spectral Density (PSD) of the thermal noise in W/Hz, $B$ is the signal bandwidth in Hz, and $N_F$ is the noise figure at the receivers.

\par For the computation of the correlation matrices $\textbf{R}_{kq},\;\forall k,\;\forall q$, we consider $N=6$ scattering clusters, $\psi_{kq,n}\sim\mathcal{U}[\phi_{kq}-40\degree,\phi_{kq}+40\degree]$, and $\sigma_{\psi}=5\degree$.

\begin{table}[]
    \centering
    \caption{Simulation parameters.}
    \begin{tabular}{l l l}
        \toprule
        \textbf{Parameter} & \textbf{Symbol} & \textbf{Value}\\  
        \midrule
        Total number of antenna elements & $M$ & 16\\
        Number of APs & $Q$ &  $\{1,2,4,8\}$\\
        Number of antennas on each AP & $S$ & $\{16,8,4,2\}$\\
        Number of active MTDs & $K$ & $10$\\
        Length of the side of the square area & $l$ & $[50,100,200]$ m\\
        Uplink transmission power & $p$ & 100 mW\\
        PSD of the noise & $N_0$ & $4\times10^{-21}$ W/Hz\\
        Signal bandwidth & $B$ & 20 MHz\\
        Noise figure & $N_F$ & 9 dB\\
        Height of the APs & $h_{\text{AP}}$ & 12 m\\
        Height of the UEs & $h_{\text{UE}}$ & 1.5 m\\
        Carrier frequency & $f_c$ & 3.5 GHz\\
        Normalized inter-antenna spacing & $\Delta$ & $0.5$\\
        Path loss exponent & $\eta$ & $2$\\
        Reference distance & $d_0$ & $1$ m\\
        \bottomrule
    \end{tabular}    
    \label{tableParameters}
    \vspace{0.3cm}
    \centering
    \caption{Parameters of the PSO Algorithm.}
    \begin{tabular}{l l l}
        \toprule
        \textbf{Symbol} & \textbf{Parameter} & Value \\
        \midrule
        $w$ & Inertial weight & $[0.1,1.1]$\\
        $c_1$ & Cognitive constant & $1.49$\\
        $c_2$ & Social constant & $1.49$ \\
        $N_{\text{vars}}$ & Number of variables & $Q$ \\
        $|\mathcal{A}|$ & Swarm size & $\min\{100,10N_{\text{vars}}\}$ \\
        $N_{\text{max}}$ & Maximum number of iterations & $200N_{\text{vars}}$ \\
        $N_{\text{stall}}$ & Maximum number of stall iterations & 20\\
        $\epsilon$ & Function tolerance & $10^{-6}$\\
        \bottomrule
    \end{tabular}    
    \label{parametersPSO}
\end{table}

\begin{figure*}
    \centering
    \begin{minipage}[t]{0.475\textwidth}
        \centering
        \begin{subfigure}[b]{1.0\textwidth}
            \centering
            \includegraphics[scale=0.58]{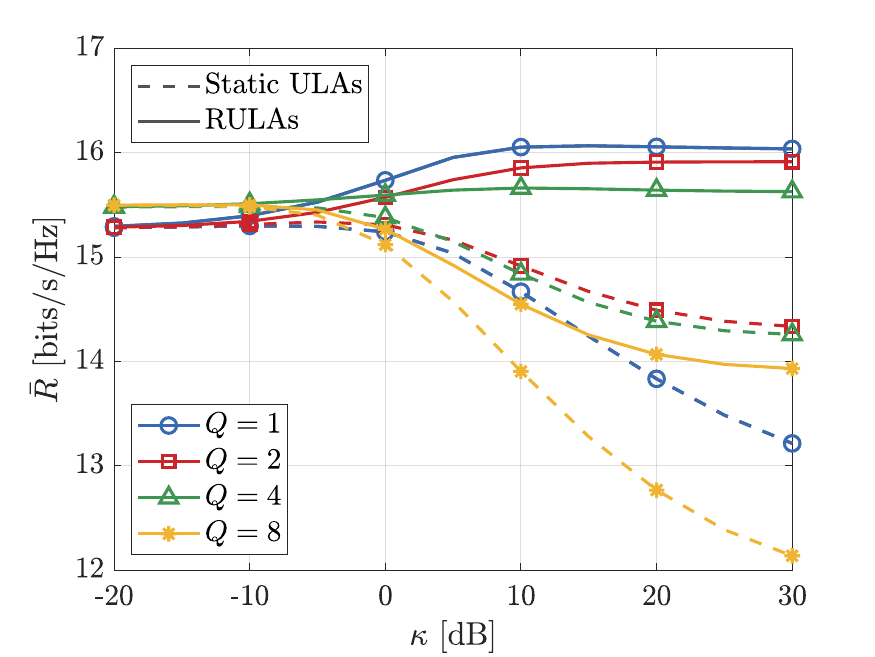}
            \caption{$l=50$~m}    
        \end{subfigure}
        \\
        \begin{subfigure}[b]{1.0\textwidth}  
            \centering 
            \includegraphics[scale=0.58]{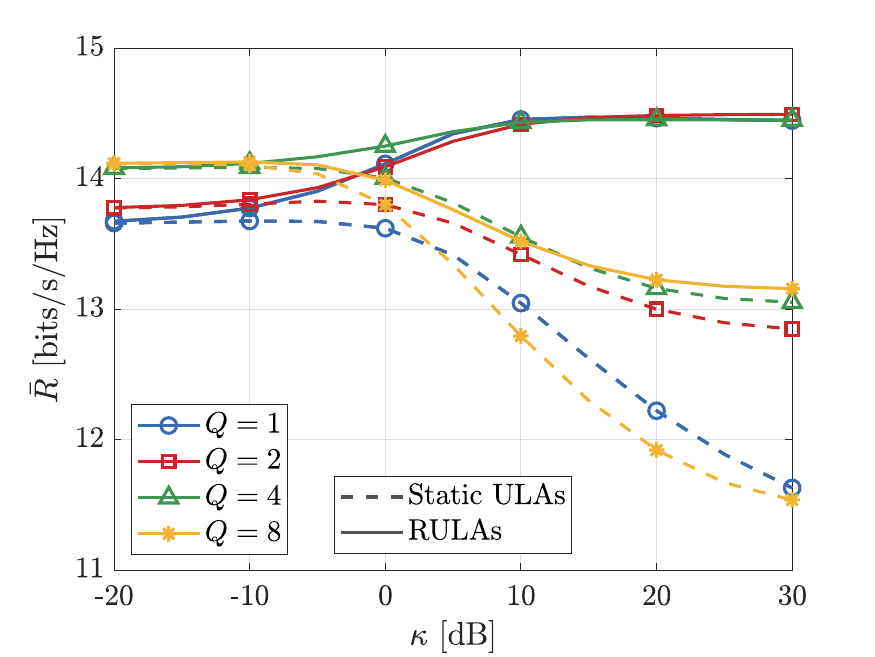}
            \caption{$l=100$~m}  
        \end{subfigure}
        \\
        \begin{subfigure}[b]{1.0\textwidth}  
            \centering 
            \includegraphics[scale=0.58]{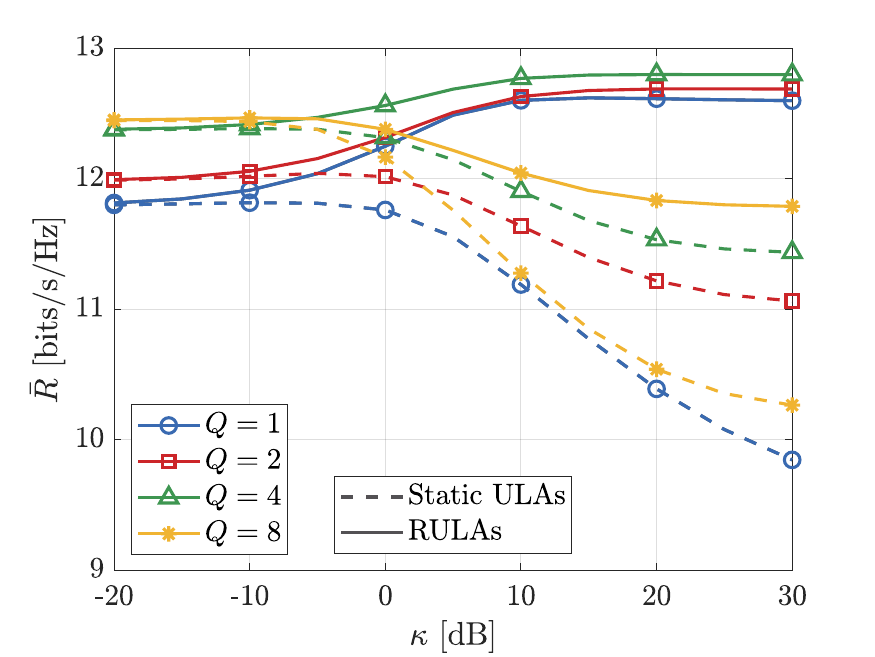}
            \caption{$l=200$~m}  
        \end{subfigure}            
        \caption{Mean per-user achievable SE versus Rician factor, considering centralized ZF and $\sigma_e^2=-10$ dB, for different sizes of the square coverage area.}
        \label{resultsKappa}
    \end{minipage}
    \hspace{0.25cm}
    \centering
    \begin{minipage}[t]{0.475\textwidth}
        \centering
        \begin{subfigure}[b]{1.0\textwidth}
            \centering
            \includegraphics[scale=0.58]{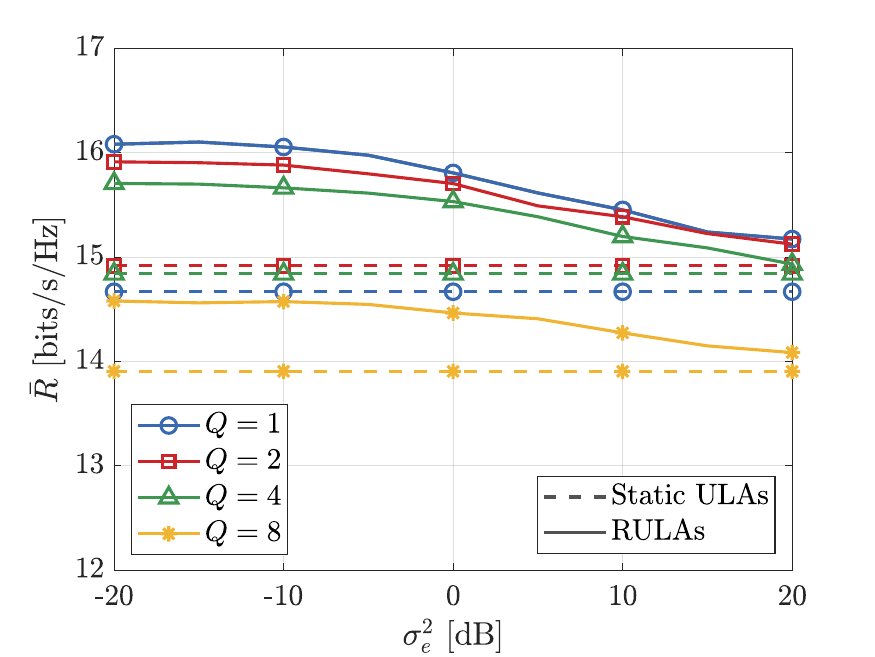}
            \caption{$l=50$~m}    
        \end{subfigure}
        \\
        \begin{subfigure}[b]{1.0\textwidth}  
            \centering 
            \includegraphics[scale=0.58]{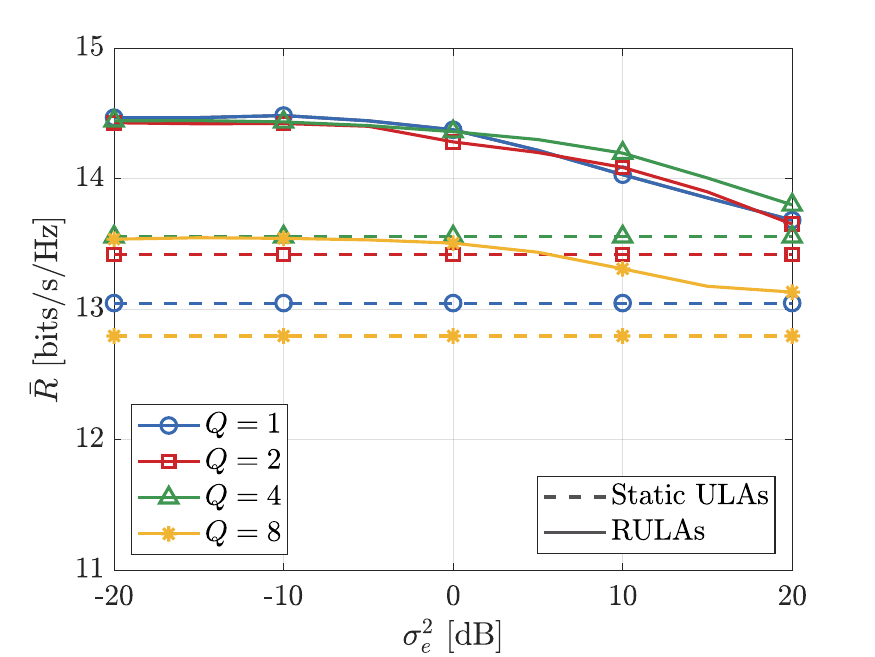}
            \caption{$l=100$~m}  
        \end{subfigure}
        \\
        \begin{subfigure}[b]{1.0\textwidth}  
            \centering 
            \includegraphics[scale=0.58]{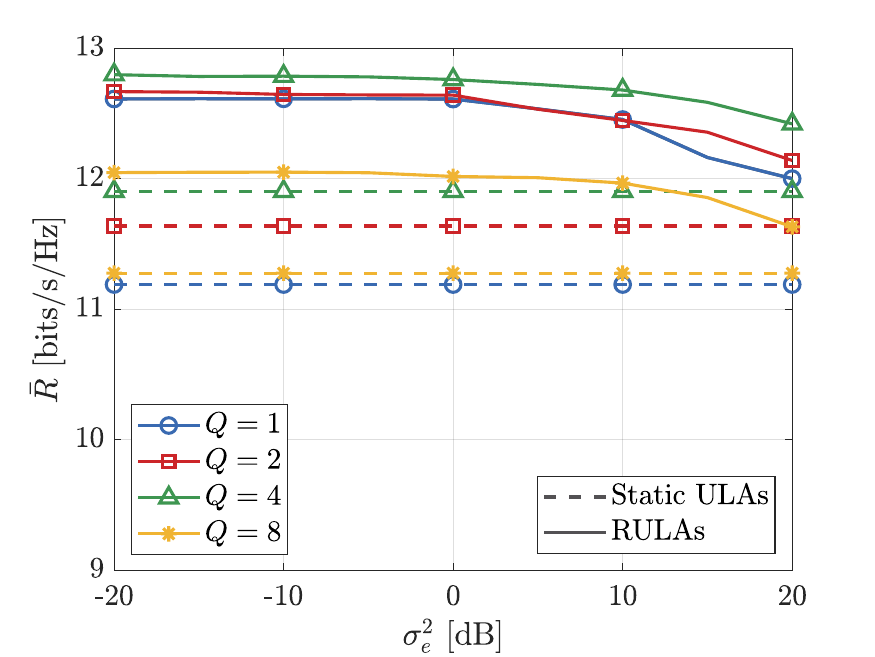}
            \caption{$l=200$~m}  
        \end{subfigure}            
        \caption{Mean per-user achievable SE versus the variance of the localization error, considering centralized~ZF~and $\kappa=10$ dB, for different sizes of the square coverage~area.}
        \label{resultsPosError}
    \end{minipage}
\end{figure*}

\subsection{Simulation Results and Discussions}

\par The mean per-user achievable SE is obtained by averaging over multiple network realizations, i.e., distinct sets of locations for the $K$ active MTDs. For each network realization, the achievable SE of the $K$ MTDs is obtained by averaging over several channel realizations, i.e., distinct realizations of the channel matrix $\textbf{H}$.

\par The mean-per user achievable SE $\Bar{R}$ as a function of the Rician factor $\kappa$, for different values of $Q$ and $l$, are shown in Fig. \ref{resultsKappa}. The primary motivation for obtaining $\Bar{R}$ as a function of $\kappa$ is to evaluate the performance of our proposed approach under a variety of different conditions. Different indoor industrial scenarios may feature diverse objects (varying in number, size and materials), the presence or absence of walls, leading to wireless channels that experience different Rician factors. We observe that, in the case of APs equipped with static ULAs, the mean per-user achievable SE decreases with the Rician factor for any values of $Q$ and $l$. This happens because the correlation among the channel vectors increases with $\kappa$, thus affecting the performance obtained with the ZF combiner. On the other hand, in the case of APs equipped with the proposed RULAs, the mean per-user achievable SE increases with $\kappa$, excepted for the setups with $Q=8$. This evinces the fact that the optimal rotation of the RULAs reduces the correlation between the channel vectors, consequently enhancing the performance obtained with ZF when the LoS component becomes strong. Note that, when $\kappa\leq-10$ dB, i.e., when the channel tends to be pure NLoS, the optimal rotation of the RULAs brings no performance improvements, as expected. Nevertheless, when $\kappa$ grows large, the performance gains obtained with the RULAs become very significant for any values of $Q$ and $l$. Moreover, when the channel tends to be pure NLoS, increasing $Q$ enhances the mean per-user achievable SE for any value of $l$. Thus, in rich scattering environments, the most distributed D-MIMO setup with $Q=8$ is always the best choice, and the setup with $Q=4$ follows closely behind. In contrast, when $\kappa$ grows large and the LoS component becomes strong, there exists an optimal number of APs (and of antenna elements per AP) that maximizes the mean per-user achievable SE.

\par The numerical results in Fig. \ref{resultsKappa} show that in the case of APs with static ULAs and under high Rician factor, the best performance is achieved by the setup with $Q=2$ in the case of $l=50$ m (Fig. \ref{resultsKappa}a), and by the setups with $Q=4$ for the cases of $l=100$ m and $l=200$ m (Figs. \ref{resultsKappa}b and \ref{resultsKappa}c, respectively). Conversely, when we adopt APs equipped with RULAs in the scenarios with high Rician factor, the single AP setup achieves the best performance for the case of $l=50$ m (Fig. \ref{resultsKappa}a). When $l=100$ m (Fig. \ref{resultsKappa}b), the setups with $Q=[1,2,4]$ achieve nearly identical performance. Finally, when $l=200$ m (Fig. \ref{resultsKappa}c), the best performance is achieved when $Q=4$. Overall, the curves in Fig. \ref{resultsKappa} show that, when the LoS component is strong, there is a sweet spot between the beamforming gains obtained with APs equipped with multiple antennas, and the macro-diversity gains obtained with the spatial distribution of APs. In the case of APs equipped with static ULAs, the intermediate deployments with $Q=[2,4]$ achieve this sweet spot. Conversely, when the proposed RULAs are adopted, utilizing a single AP allows us to achieve satisfactory performance while avoiding D-MIMO networks' higher deployment and maintenance costs.

\par In Fig. \ref{resultsPosError}, we show the mean per-user achievable SE versus the variance of the localization error $\sigma_e^2$, for different values of $Q$ and $l$, and considering $\kappa=10$ dB. Note that all the curves in Fig. \ref{resultsPosError} corresponding to the case of static APs are constant since those setups do not use the estimated locations of the active MTDs for any optimization. As discussed in Section \ref{localizationErrorModel}, the scenarios where $\sigma_e^2\geq5$~dB, that is, with very poor location estimates, are not expected to occur in current or future wireless networks under normal operational conditions. We evaluate this range of variance of the localization error to illustrate the robustness of our proposed framework. Interestingly, we observe that the performance gains owing to the optimal rotations of the RULAs are still very significant even when the accuracy of the location estimates is poor. Nevertheless, the SE is almost constant for $\sigma_e^2\leq-10$ dB, that is, improving the accuracy of the location estimates does not yield noticeable performance improvements. The curves in Fig. \ref{resultsPosError}a show that, in the case of a small coverage area (specifically, $l=50$~m), it is more advantageous to deploy fewer APs equipped with more antennas since the best performance is achieved with $Q=2$ in the case of static ULAs and with $Q=1$ in the case of the proposed RULAs. When we increase the size of the coverage area to $l=100$~m and $l=200$~m (Figs. \ref{resultsPosError}b and \ref{resultsPosError}c, respectively), we observe that the setup with $Q=4$ achieves the best performance for both cases of static ULAs and RULAs. In the case of $l=100$~m, the performance achieved with $Q=1$ and $Q=2$ when RULAs are adopted approach the performance obtained with $Q=4$. These numerical results show that, as we increase the size of the coverage area, and given a total number of antenna elements $M$, a sweet spot between the number of APs and the number of antenna elements per AP becomes evident. This sweet spot corresponds to a trade-off between macro-diversity and beamforming gains. Finally, it is also very interesting to observe that in all the cases ($l=\{50,100,200\}$~m), the most distributed setup with $Q=8$ presents the worst performance.

\subsection{Complexity Analysis of PSO}

\par The simulations ran on a personal computer having Windows 10, Intel(R) Core(TM) i5-10210U processor and 32 GB of RAM memory. Besides, we utilized Matlab version R2022b. The PSO was implemented on Matlab using the \texttt{particleswarm} built-in function \cite{mathworks_particleswarm}. Since we utilized a built-in function of Matlab, it is very difficult to measure the complexity of the PSO in terms of metrics such as number of float-point operations (FLOPs). Thus, we decided to analyse the computational complexity of our PSO solution in terms of the mean and the variance of the required time for the optimization algorithm to finish. This time can be measured in Matlab using the functions \texttt{tic;toc} or \texttt{cputime}.

\begin{table}[t]
    \centering
    \begin{tabular}{l l l l l}
        \toprule
        \makecell{Number of\\RULAs} & $Q=1$ & $Q=2$ & $Q=4$ & $Q=8$\\
        \midrule
        Mean & 0.2652 s & 0.6042 s & 2.1520 s & 7.9548 s\\
        \midrule
        Variance & 0.0363 s$^2$ & 0.0636 s$^2$ & 0.4624 s$^2$ & 7.3269 s$^2$\\
        \bottomrule
    \end{tabular}
    \caption{Mean and variance of the execution time required for PSO to finish, considering $\kappa=10$~dB, $\sigma_e^2=-10$~dB and $l=100$~m.}
    \label{tableTime}
\end{table}

\par The mean and variance of the execution time can be found in Table \ref{tableTime}. The PSO converges when the relative change in the value of the objective function over a given number of ``maximum stall iterations'' is less than the ``function tolerance'' \cite{mathworks_particleswarm}. We observe that the mean required time increases rapidly with the number of RULAs $Q$ (i.e., the number of optimization variables). Therefore, PSO indeed may not be the best choice in D-MIMO networks with multiple RULAs whose angular positions are jointly optimized by the CPU. Moreover, the mean required time for any given value of $Q$ indicates that it is impractical to optimize the angular positions of the RULAs at every time slot. In other words, the angular positions should not be adjusted too often. In the framework we propose, the RULAs' angular positions are adjusted a single time, prior to the uplink data transmission phase. Subsequently, the angular positions remain unchanged throughout multiple time slots during the uplink data transmission phase.

\section{Conclusions}
\label{conclusions}

\par In this work, we studied the performance of D-MIMO networks for indoor scenarios. Aiming to enhance the quality of the wireless links, we proposed the deployment of APs equipped with RULAs. The CPU jointly computes the optimal rotation of the RULAs as a function of the estimated locations of the active MTDs. The optimization is done using a location-based beamforming approach and PSO. We considered a spatially correlated Rician fading model and evaluated the setups' performance for different Rician factor values. We took into account the impact of imperfect localization information and also of the imperfect CSI. Given the total number of antenna elements, we evaluated the trade-off between the number of APs and antennas per AP. The numerical results show that the optimal rotation of the RULAs can bring substantial performance gains in terms of mean per-user achievable SE, and the gains grow with the Rician factor. We also observed that the optimal rotation of the RULAs can bring performance gains even when the localization accuracy is poor or moderate. We also concluded that, given the total number of antenna elements, there is a sweet spot between the number of APs and the number of antenna elements per AP, corresponding to a trade-off between beamforming and macro-diversity gains. The best performance is achieved by adopting a few APs equipped with multiple antennas. This approach also has the advantage of reducing the deployment and maintenance costs of the system and its computational complexity.

\par This work studied the performance gains in terms of SE that the rotation of the ULAs can bring to wireless communications. Nevertheless, Integrated Sensing and Communications (ISAC) is a research topic that has become extremely popular in both academia and industry. Some recent works \cite{ma2024,lyu2024,khalili2024} have evaluated the performance improvements in the accuracy of sensing and localization obtained with the movement of antennas. The movement of antennas can be beneficial in several ways. For instance, it can help overcome blockages in the line-of-sight between an access point (AP) and a device. Additionally, by allowing for precise beamforming and optimal beamwidth design, movable antennas can enhance data transmission and improve sensing accuracy. This capability reduces undesirable side lobes and improves interference management. Furthermore, movable antennas can dynamically switch between optimal array geometries, depending on whether they are used for sensing or communication applications. Comparing different implementations of movable and/or rotary antennas and the associated performance improvements for both sensing and communications is a promising direction for future works.

\appendices

\section{Parameters to Ensure Far-Field Propagation}
\label{farFieldPropagation}

\par The Fraunhofer distance determines the threshold between the near-field and far-field propagation and is given by \cite{sherman1962}
\begin{equation}
    d_F=\dfrac{2D^2}{\lambda},
\end{equation}
where $D$ is the largest dimension of the antenna array, $\lambda=c/f_c$ is the wavelength, $c$ is the speed of the light and $f_c$ is the carrier frequency. In the case of an RULA with half-wavelength spaced antenna elements, we have
\begin{equation}
    D_{\text{RULA}}=(S-1)\lambda/2,
\end{equation}
where $S$ is the number of antenna elements of the RULA. Considering $f_c=3.5$ GHz, the height difference between the APs and the MTDs must be at least 10 m to ensure far-field propagation conditions for all the devices. Thus, in this paper we set $\max\{S\}=M=16$ and $h_{\text{AP}}=12$ m.

\bibliographystyle{./bibliography/IEEEtran}
\bibliography{./bibliography/references}

\end{document}